# Pathway Toward a Mid-Infrared Interferometer for the Direct Characterization of Exoplanets.

Jean Schneider
Paris Observatory
Jean.Schneider@obspm.fr

## Abstract
We recognize the need for the characterization of exoplanets in reflected light in the visible and in the IR termal emission.
But for the thermal infrared we also recognize the difficulty of an interferometric nuller
We nevertheless endorse the need for future interferometers.
We propose a new, realistic, pathway to satisfy both goals, thermal infrared studies and interferometric architectures.

## Introduction
For completeness and clarity, let us remind that the ultimate goal of exoplanet science is, in addition to comparative exoplanetology, the search for Life in the Universe.
On this pathway, we need:
1. a flexible notion on what is meant by « Life »
2. a flexible view of what is a potentially habitable planet (i.e. a planet where Life can develop)
3. a statistics on the frequency of potentially habitable planets
4. the identification of a minimum number (a few tens) of potentially habitable planets
5. the identification of biomarkers, mode of observation, wavelength range and spectral resolution of detectability
6. the identification of ground based facilities and space missions suited for these observations.

Point 3 is almost done and point 4 is under way.

## The paradigm « thermal IR-interferometer/visible-coronagraph »
NASA has organized in 2000-2002 a series of competitive studies for TPF architectures.
See http://planetquest.jpl.nasa.gov/TPF/arc_index.cfm .
For obvious reasons, JPL had to satisfy the two main communities, namely the single aperture approach and the interferometric approach. On the other hand, transversally to the instrumental aspects, from the scientific objectives point of view two wavelength regimes are to be considered: visible reflected light and infrared thermal emission. They are complementary, each one having its specific advantages. Quite naturally, the simplest (most compact) mono-aperture approach selected by JPL was a coronagraph in the visible, or TPF-C. Mechanically, the interferometric approach was thus selected for the infrared emission regime, namely TPF-I. We partly still live with this heritage.

## An invitable need for the long term: interferometry.
After the first low resolution spectro-imaging of exoplanets, a first obvious step will be to increase the spectral resolution and the sensitivity of space missions. But, sooner or later, higher angular resolution will also be needed. Indeed, after the image of planets as single pixels, the imaging of planetary rings and moons will be the next step. The morphology of exo-rings, combined with the architecture of exo-moon systems, will help to understand their formation and dynamics (in the Solar System, the origin of Saturn rings seems still problematic). Moons of giant planets in the habitable zone are also potential sites for life. And later on, in a far but not unrealistic future (Labeyrie 1996, Labeyrie et al. 2008, Schneider et al. 2005), it will be necessary the make the multipixel cartography of planet surfaces. They will be necessary to understand the exact ocean/continents morphology of planets, importance of polar caps, geographic repartition of « exo-vegetation » etc. In addition to exoplanetology, general astrophysics will of course benefit from very high angular resolution. Above a certain angular resolution, very long baselines, and therefore interferometers, will become inevitable, even in the visible (Schneider 2004).

## A bottleneck in the interferometric approach
Unfortunately, the first step of the interferometric approach, i.e. Bracewell-like nulling interferometry with a few apertures, presents some difficult aspects:
– it requires high contrast/nulling performances (like any other approach)
– the choosen nulling mechanism (Bracewell nulling) allows only a mono-pixel observation of the whole planetary system (with the subsequent problem of the background of exo-zodis)
– it requires high performance free flyers metrology
– by definition the sources (exoplanets) are extremely faint and contain only a few photons during the time scale of stability of the system (« instability noise »).

By themselves, each of these difficulties are not unsurmontable. The difficulty comes from their accumulation at the same time.
One could imagine to start with a precursor, like FKSI. But the problem is that there seems to be no simple precursor
– technologically « easy »
– scientifically exciting

We face a bottleneck.
How to escape it? How to build an interferometric mission and satisfy the thermal infrared approach?

## A prerequisite: breaking the « Earth's twin » paradigm. « Earth's twins » are the wrong targets.
« Earth's twins » do essentially retain the attention of journalists and the general public on sentimental grounds and would constitute a new geo-centrism. But they are the most difficult to detect. It is not rational to start with the most difficult targets. Let's rather start with targets easier to detect but not less interesting, i.e. Super-Earths. Super-Earths ($2R\_Earth$) are as well interesting since all terrestrial concepts (solid/liquid surface, plate tectonics, habitability) apply to these planets. See the White Paper ''Diversity among other worlds: characterization of exoplanets by direct detection'' astro-ph/0811.2496 for more details.

## Breaking the deadlock by changing paradigm: A two step solution.
Here we present an approach which decouples the needs for a thermal infrared approach and for future interferometers:
– A large IR coronagraph for close-by Super-Earths
– An interferometric precursor scientifically valuable and technically feasible

## A large IR coronagraph for close-by Super-Earths
Contrary to an interferometer with a few pupils, a large coronagraphic single aperture presents two main advantages:
– it provides a full 2D image of the planetary systems instead of a single pixel signal, making the exo-zodi problem less severe. Other single aperture like external occulters or Fresnel imager are proposed, but they are better suited for the UV/Visible approach.
– For the same angular resolution, the collecting area is at least 10 times larger than for the TPF-I architecture.

*Science requirements, instrument characteristics and expected performances*
– We start with the following scientific requirements :
    – Detect and characterize at least 10 super-Earths near the habitable zone HZ at

around 1 AU of parent star. Mayor et al 2008 have recently estimated that at least 30% of KGF stars have super-Earth with M < 25 M_Earth and periods < 25 days. They also estimate that 80% of these planets are in multiple systems, leading to the reasonable hypothesis that in these systems there is often a planet near the HZ. By applying a **« factor of security » of 2,** we thus require that the instrument can observe $10/(30\%/2) = 70$ stars.

These 70 stars will not be searched by the mission, but will be previously selected by Radial Velocity surveys.

- Detect the spectral features of the most interesting species, i.e. H2O at 6-7 micron, O3 at 9.6 micron and CO2 at 14-18 micron. In addition, it is desirable the detect the central inversion peak of the CO2 band to evaluate the vertical temperature distribution of the planet. Otherwise, the interpretation of data would be unclear.

- The science requirements translate in the following specification for the telescope and the spectrograph:
    - Telescope diameter: the 70 nearest stars lie up to approximately 4 pc, corresponding to a mimimum angular separation star-planet of 250 mas. Assuming that the coronagraph can characterize planets down to an angular distance of $1.5\lambda/D$ at 18 micron (see below) for the farest stars ($D$ = telescope diameter), this leads to a diameter of 20 m
    - Spectral resolution: R = 20 to detect the central inversion peak of the CO2 band at 15 micron (see Figure).

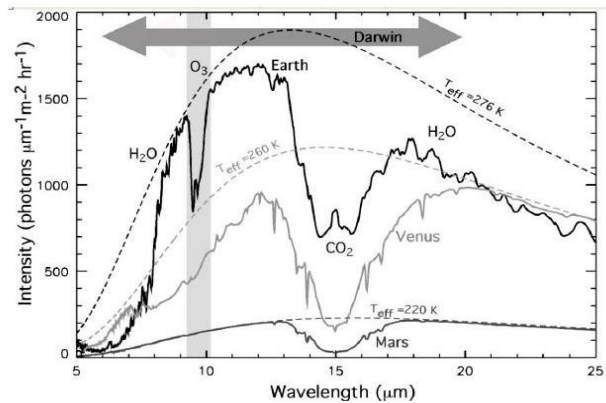

With these characteristics, ($D$ = 20 m, R = 20), the performances of the instrument will be the following:
- Ideal optics without wavefront errors, without exozodi background.
    In that case the SNR is the photon noise from the planet alone.
    For R = 20, an exposure time of 10 hours and a 2 R_Earth planet at 4 pc we have:
    - SNR = 100 in 1 hour for the integrated 6-18 micron spectral domain
    - SNR = 20 in 1 hour per spectral channel (R=20)
- In presence of wavefront errors, the planet is embeded in stellar speckles. The planet/speckle ratio depends on the coronagraphic mask performances, the amplitude of wavefront errors of the primary mirror, the performance of suppression of wavefront errors and the calibration of speckles .
    Without entering here into details, many high performance coronagraphs are presently studied: phase mask coronagraphs, pupil-shaped coronagraphs, phase-induced amplitude apodizer (PIAA, Martinache et al. 2006) and their combinations. They have been proven to work also in the mid-Infrared (Baudoz et al. 2006). The suppression of wavefront errors can be made by deformable mirrors (e.g. Trauger & Traub 2007). The wavefront corrections and speckle calibration can be made e.g. by a « self-coherent camera » (Galicher et al 2008). They should be adaptable to the thermal infrared.
    For instance, by extrapolating to a 20 m telescope with wavefront errors of 200 nm (diffraction limited mirror) the results of simulations (with a « Self Coherent Camera » (after full post-processing) by Galicher et al (2008 see Fig. 2), we find

- SNR = 5 in 1 hour for the integrated 6-18 micron spectral domain
- SNR = 5 in 4 hour per spectral channel (R=20)

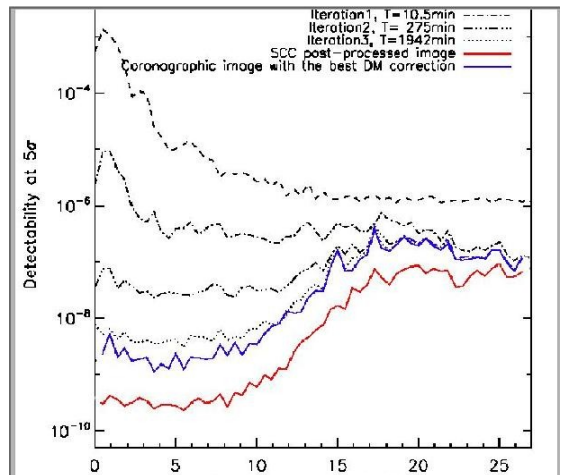
Figure 2:

*Payload and instrument*
- Telescope. Two options:
    - A deplyable prime mirror. In order to minimize the wavefront imperfections due to the adjustement of several single pupils, we suggest to have a mirror in 2 elemnts shaped as semi-circle each with a diameter of 20 m  The overall dimension of each pieces would be 10 m x 20 m, fitting in the ARES V fairing with a diameter of 10 m. Four 10 m elements  are also possible.
    - A single elliptical or rectangular mirror with dimensions  10 m x 20 m. Assuming that in 2025 we will know the orientation in space of planetary orbits, the larger dimension can be oriented in order to have the desired angular resolution along the star-planet vector. In that case, the SNR degrades only by a factor of $\sqrt{2}$ compared to a circular 20 m aperture.
- High contrast imaging. Its performanes rests on three aspects: PSF quality, wave front calibration and control, and coronagraphic rejection factor.
    - The PSF quality should be comparable to or better than the EPICS exoplanet camera at the E-ELT PSF which has currently 906 elements
    - Wave front control: deformable mirrors
    - Corongraph: PIAA, phase mask, shaped pupil and combinations.

Such an instruments has been extensively studied by the former TRW Company (presently Nothrop Grumman) and submitted to JPL. See the study results at
http://planetquest.jpl.nasa.gov/TPF/TPFrevue/FinlReps/Trw/TRW12Fnl.pdf
One of the scientific reasons for not being selected was the large « inner working angle » IWA ( $4-5\lambda/D$ ) due to the choice of a pure Lyot stop. With current phase masks, the IWA can be much lower.
For other details like deployment, cryogenics, solar shielding etc, see the TRW study.

## A ground-based precursor: METIS at the E-ELT
METIS, the Mid-infrared ELT Imager and Spectrograph, is a proposed instrument for the European Extremely Large Telescope (E-ELT), currently undergoing a phase-A study (Brandl et al 2008). The study is carried out within the framework of the ESO-sponsored E-ELT instrumentation studies. METIS will be designed to cover the E-ELT science needs at wavelengths longward of 3um, where the thermal background requires different operating schemes. Exoplanets are among the main science drivers from which the instrument baseline has been derived. Specific emphasis has been given to observations that require very high spatial and spectral resolution, which can only be achieved with a ground-based ELT. The challenging aspects of background suppression techniques, adaptive optics in the mid-IR, and telescope site considerations are presently under study. The METIS instrument baseline includes imaging and spectroscopy at the atmospheric L, M, and N bands with a possible extension to Q band imaging. Both coronagraphy and polarimetry are also being considered.

But then the problem of having a valuable precursor for an interfereometer still remains. This leads us to the second step of our proposed plan.

## A space interferometric precursor scientifically valuable and technically feasible

Instead of accumulating problems due to free-flying control, faintness of exoplanets, single pixel detection, high contrast, we propose to have a simpler instrument with nevertheless first class science:
– no nulling/high contrast
– bright targets
– no a priori wavelength constraints due to angular resolution (they are technically relaxed for free-flyers)
– different science: general astrophysics, not mainly exoplanets. For instance:
    – size and flattening of interesting stars
    – perturbation of star (apparent) position by passing gravitational waves
    – astrometric perturbation of stellar centroid by transiting planets
    – lensing (multiple quasars, « stellar arcs », astrometry of planetary lensing...)
    – super hot super-Earths

The idea is to start with 2 apertures, free-flying or not.

**By this double approach the scientific objectives (characterization of exoplanets in the thermal IR) of the TPF-I community and the need for a precursor to interferometers with high contrast imaging will be both satisfied.**

## Resulting « roadmap »

We therefore propose a roadmap which would start with a 2m class coronagraph in the visible for targets previously detected by radial velocity, followed by mid-IR instruments and large interferometers. In short, successively:
– Intense Radial Velocity survey to detect super-Earths near the habitable zone.
– A 1.5-2 m visible coronagraph
– A ground-based Mid Infrared coronagraph at the E-ELT (METIS)
– A 20 m space coronagraph in the Mid Infrared
– A simple 2 aperture interferometer (without nulling) for very high angular resolution general astrophysics and exoplanet science on bright sources
– Multi-aperture interferometers with a coronagraph or nuller

## Technological heritage, technologies under development and to be developped

HST --> JWST: x3 + deployment technology + cryogenics
JWST --> Large Mid-IR Coronagraph: x3
– in orbit deployment of segmented mirror: heritage from JWST
– large fairing launcher (10 m wide or more): under developpement (ARES V, first flight: 2018)
– Mid IR coronagraphic masks: heritage from JWST
– Cryogenics: heritage from JWST
– High quality optics prime mirrors (Wave front errors 200 – 400 nm): technology to be developped for 10m mirror. HST heritage: WFE = 20 nm for 2.4 m mirror. SPICA heritage:

- 350 nm for 3.4 m mirror.
- In case of segmented prime mirror, high quality PSF: heritage of E-ELT
- Cryogenic deformable mirror: SPICA heritage (under development)